Authors: Nicholas Cummins[1,2,3,4], Lauren L. White[1], Zahia Rahman[1], Catriona Lucas[1], Tian Pan[1], Ewan Carr[1], Faith Matcham[5], Johnny Downs[2], Richard J. Dobson[1,3,6], Thomas F. Quatieri[7], *Judith Dineley[1,2]

1. Department of Biostatistics & Health Informatics, Institute of Psychiatry, Psychology & Neuroscience, King's College London, London, UK,
2. CAMHS Digital Lab, Department of Child and Adolescent Psychiatry, Institute of Psychiatry, Psychology & Neuroscience, King's College London, London, UK
3. King's Institute for Artificial Intelligence, King's College London, London, UK
4. thymia, London, UK
5. School of Psychology, University of Sussex, Falmer, UK
6. Institute of Health Informatics, University College London, London, UK
7. MIT Lincoln Laboratory, Lexington, MA, USA
* Corresponding Author: judith.dineley@kcl.ac.uk

# A methodological framework and exemplar protocol for the collection and analysis of repeated speech samples

## Abstract

**Background:** Speech and language biomarkers have the potential to be regular, objective assessments of symptom severity in several neurological and mental health conditions, both in-clinic and remotely using mobile devices. However, the complex nature of speech and often subtle changes associated with health mean that findings are highly dependent on methodological and cohort choices. These are often not reported adequately in studies investigating speech-based health assessment, which (i) hinders the progress of methodological speech research, (ii) prevents replication, and (iii) makes the definitive identification of robust biomarkers problematic.

**Objective:** 1. To facilitate replicable speech research by presenting an adaptable speech collection and analytical method and design checklist for other researchers to adapt for their own experiments.
2. To develop and apply an exemplar protocol that reduces and controls for confounding factors in repeated recordings of healthy speech, including device choice, speech elicitation task and non-pathological variability.

**Methods:** We developed a collection protocol based on a thematic literature review. Our protocol comprises the elicitation of (1) read speech, (2) held vowels, and (3) a picture description. With a focus towards remote applications, we collected speech with different devices: a freestanding condenser microphone, three smartphones and a headset. We developed and report in detail a pipeline to extract a set of 14 exemplar speech features, also chosen via a thematic literature review, that cover timing, prosodic, quality, articulatory, and spectral characteristics of speech.

**Results:** We collected healthy speech from 28 individuals three times in one day (*Day*), repeated at the same times 8-11 weeks later, and from 25 individuals on three days in one week at fixed times (*Week*). Participant characteristics collected included sex, age, native language status, and the participant's voice use habits. Before each recording, we collected information on recent voice use, food and drink intake, and emotional state. Recording times were also documented. Values of the extracted features are presented providing a resource of normative values.

**Conclusions:** Speech data collection, processing, analysis and reporting towards clinical research and practice varies widely, motivating this report and speech collection protocol design checklist. Greater harmonization of study protocols and consistent reporting are urgently required to translate speech processing into clinical research and practice.

## Introduction

The linguistic and paralinguistic content of our speech contains rich information on our cognitive, neuromuscular and respiratory functioning. There is a growing body of literature highlighting the potential of speech as an objective diagnostic, monitoring and predictive marker in a variety of clinical cohorts including amyotrophic lateral sclerosis [1], Parkinson's disease [2, 3, 4], psychosis [5, 6] and major depressive disorder (MDD) [7, 8] as summarized in several reviews [9-11]. Key advantages of recording speech for clinical applications include that it is non-invasive and can be conducted, both in the clinic and remotely, using off-the-shelf consumer-grade audio equipment.

However, accurately detecting changes in speech driven by changes in health is challenging, and speech markers are yet to be used as an outcome measure in clinical trials or translated for clinical use. This is partly because speech is a multi-faceted, complex, dynamic signal. Many speech changes associated with different health states can be subtle, forming one part of a measured signal that is also dictated by other speaker-specific factors and recording and analysis choices. There is a pressing need to understand, quantify, and adjust for the effect of such variables as they can mask or even mimic the effect of health changes.

Potentially confounding speaker factors include hormonal variations within the menstrual cycle [12, 13], fatigue [14], voice use habits [15-17], emotion [18] and hydration [19]. Systematic changes with age, menopause, and medication use have also been reported [20-22]. A growing body of literature highlights the impact of methodological choices, including recording environment, hardware choices, digitization formats, and choice of extraction tools on speech characteristics and subsequent health state analysis [23-28]. Speech elicitation strategies are another important factor in speech-based health assessment. Common strategies include reading passages, image description tasks, free response questions, and vocal function exercises such as sustained phonation [10, 11]. Each task can produce distinct acoustic, linguistic and emotional content, so choosing the right tasks is vital for ensuring the clinical validity and sensitivity of the extracted speech measures [29].

Practice effects represent another potential confounder, where recorded speech changes due to repeated exposure to a task or activities [30, 31]. Though expected in speech research, practice effects are rarely documented [9, 32].

Despite an awareness of such effects, methodological details and important speaker characteristics in speech-based health assessment research are under-reported in the literature. These factors can be unaccounted-for sources of variation that become particularly pertinent when effect sizes are small or context-dependent [33], which is often the case in health analyses of speech. This is of particular concern for remote data collection outside of laboratory settings where there are more degrees of freedom, e.g., recording devices and geometry and the acoustic environment. This weakens replicability and hinders the development of robust methodology and tools, the discovery and verification of biomarkers and, ultimately, clinical translation.

The lack of established methods for data collection and reporting exacerbates these issues [34]. The Consensus Auditory-Perceptual Evaluation of Voice (Cape V) protocol [35] and recommendations made by the American Speech-Language-Hearing Association (ASLHA) Panel [36] are helpful starting points. However, these recommendations were developed for in-lab speech pathology assessments. They have limited applicability in detecting subtle changes and the broader range of speech characteristics associated with, for example, mental health and neurological disorders recorded remotely and longitudinally.

The Voiceome Study represents an attempt at standardization of longitudinal data collection for speech and language biomarker research [37]. A key feature is its recommendation of 12 speech elicitation tasks, and the authors highlight these tasks produce distinct feature clusters. However, the authors do not describe the clinical relevance of these prompts or provide any evidence base justifying their inclusion. The implications for participant burden and associated protocol acceptance and adherence by participants are also not discussed, which is an important issue in data collection [38]. The effects of the recording environment, recording time, hardware choices, and speech processing methods on the quality of extracted data are also not considered.

In conclusion, the effect of speaker factors and methodological choices necessitates harmonizing data collection and reporting methods across speech and language biomarker research. This step will aid transparency and reproducibility, enabling the cross-study interpretability of key findings and insights.

## The King's College London Voice and Speech Lab Protocol for minimizing methodological between-recording variability

We developed the presented protocol as part of a pilot study whose aim was to address some of the challenges previously outlined above and contribute new knowledge on non-pathological variability in speech production over repeated recordings. The consideration of participant burden and the acceptability of the protocol to participants was part of this process, as these factors have important implications for recruitment and protocol adherence, and therefore data quality and completeness. While our protocol was designed with a specific scientific goal in mind, the core methodological aspects cover design and reporting decisions relevant for researchers collecting longitudinal data for speech and language biomarker research. Such research will benefit from minimizing between-recording variations in recording speech due to methodological factors and clear reporting of methodology. By presenting our methodological choices, other researchers can adapt our protocol to address their own research questions, thereby facilitating replicable speech research.

Specifically, the protocol was developed to collect data for the assessment of within-individual variability in speech over a single day and a single week while minimizing variability due to other factors. With remote measurement and digital health applications in mind, we recorded speech with several devices, including smartphones, to observe the capacity of different devices to capture speech variability. As a pilot investigation, we collected and analyzed speech samples from healthy participants to avoid variability driven by pathology. Datasets of healthy individuals are also beneficial as baselines for comparison with clinical populations in both speech pathology and neurological and mental health assessment [29].

## Protocols for investigating within-individual speech variability

Most protocols in the literature have been part of studies assessing localized vocal tract pathology and dysphonia, analyzing a small number of speech characteristics relevant to localized speech pathology, typically with modest-sized cohorts. Many of these studies were also conducted before remote recording and mobile devices were a consideration [39, 40].

Most recently, however, Pierce et al., trained female participants in a remote recording study [41]. Participants completed one baseline supervised recording and then recorded themselves three times each day for a week within prescribed intervals in a well-described protocol. The 45 participants read aloud two texts and performed sustained vowels each time using a cardioid head-mounted microphone. Participants were advised to record in a quiet room with no tiling; however, adherence to this was not reported. Pierce et al. analyzed 32 speech features. However, typical of studies motivated towards conventional speech pathology assessment, timing characteristics, which provide insights into several neurological and mental health conditions, were not investigated [41].

Several studies motivated by mental and neurological disorder assessment have quantified within-speaker change, framed as test-retest reliability assessment. Feng et al. recorded 40 healthy young adults twice 2-3 days apart in the same test room, completing seven elicitation tasks in Mandarin [42]. Barnett et al. retrospectively analyzed speech features of 46 healthy individuals recorded twice, months apart, reading aloud the Bamboo Passage [43]. Stegmann et al. reported an analysis of 22 healthy individuals recorded daily for seven days, and clinical cohorts with amyotrophic lateral sclerosis (ALS) (72 participants) and frontotemporal dementia (24 participants) on recorded approximately a week apart [44].

However, in each of these analyses, various methodological details such as consistency in recording time and acoustic conditions – and adherence to instructions in the unsupervised ('in-the-wild') recordings – are not reported. At least in principle, therefore, measurement factors may be responsible for a proportion of the observed differences between repeated recordings of a given participant. An additional potential limitation of these works is the use of the same elicitation scripts in each recording. Increased speaker familiarity with the readings can result in practice effects [30, 31], which could confound the assessment of within-individual speech variability [9, 32]. Finally, to the best of the author's knowledge, none of the aforementioned studies have provided data (either raw audio or extracted features).

Comparing key methodological choices, our protocol improves on these previous works in that we collected data at set times, in a controlled, supervised environment and used multiple microphone types (Table 1). We also provide a rich and varied set of elicitation tasks. section can include background information such as theories, prior work, and hypotheses.

Table 1. Comparison of key methodological choices in protocols of studies observing within- and between-speaker variability.

| Study | n | Cohort type | Schedule | L v R[a] | Microphone | Speech type[b] |
|---|---|---|---|---|---|---|
| Cummins et al, 202X | 28 | healthy | 3/day, twice in 8-11 wks. fixed times | L | condenser 3 phones 1 headset | R, SV, PD |
| Cummins et al, 202X | 26 | healthy | 3 in 1 wk. fixed days fixed time | L | condenser 3 phones 1 headset | R, SV, PD |
| Garrett & Healey, 1987 [39] | 20 | healthy | 3 in 1 day | L | miniature condenser | R |
| Leong et al, 2013 [40] | 18 | healthy | 10 in 30 days, fixed time interval | L | moving coil | R, SV |
| Pierce et al, 2021 [41] | 45 | healthy | 3/day, 1 wk. | R | headset condenser | R, SV |
| Barnett et al, 2020 [43] | 46 | healthy | 2 in 3-6 mos. | NES[c] | NES | R |
| Stegmann et al, 2020 [44] | 72 | ALS | daily, 1 wk. | R | NES | R, SV |
|  | 22 | healthy | daily, 1 wk. | R | NES | R, SV |
|  | 24 | ALS & dementia | 2 in ~1 wk. | NES | NES | R, SV, PD |
| Feng et al, 2024 [42] | 40 | healthy | 2 in 2-3 days | L | condenser | R, SV, CS, RS, DDK |

[a] Recording location, L: Laboratory, R: Remote
[b] Speech elicitation types: Sc: Scripted, SV: Sustained vowels, CS: Connected speech, DDK: diadokinetic rate test, RS: repetition of heard speech
[c] NES: Note explicitly stated by authors

## Methods

Herein, we describe our protocol for capturing repeated speech samples with minimized measurement variability. We describe multiple methodological details relevant to wider speech and language biomarker research. To facilitate adaptation to new protocols addressing other research questions, we provide a checklist of key considerations (Multimedia Appendix 1).

This protocol was designed for a pilot study whose primary focus was to assess within-speaker non-pathological variation in speech over time. In Day, we aimed to record healthy volunteers speaking (a) in the morning, afternoon, and early evening of a single day (Day 1), and (b) repeated at the same times on a second day 8-11 weeks later (Day 2). In Week, our aim was to record healthy volunteers on three days in one week at the same time each day. The pilot study received approval from the Research Ethics Committee of King's College London (reference: LRS/DP-22/23-36194).

## Recruitment

Adult staff and students at the study institute and local residents were recruited via advertisements in a research recruitment circular, institute email lists, on social media and with physical flyers and posters. Potential participants were asked to read an information sheet and complete a pre-enrolment screening form that repeated the eligibility criteria and collected contact details and sociodemographic data to facilitate the recruitment of a balanced cohort.

We excluded individuals under 16 or over 65; over-65s were excluded to minimize speech effects associated with old age [45]. We also excluded smokers, those with dyslexia, and individuals currently receiving treatment for any speech, auditory, mental, neurological, respiratory, or other health disorder that could impact their speech. Additionally, we excluded non-native English speakers unless they had a sufficient level of English proficiency to read an intermediate or advanced text aloud.

We regularly checked the cohort balance throughout recruitment to enable timely, targeted recruitment as needed. Sociodemographic groups that were under-represented at pre-enrolment – male participants and participants over 30 – were prioritized for follow-up and recruitment. After an initial round of advertising, in subsequent advertising, we advertised for male participants exclusively.

Researchers emailed individuals to allocate them to three recording sessions in one day (Day) or week (Week) according to their availability and preference. Emails at each stage of participation used text templates individually adapted for more personable communication to encourage engagement. Each provisional participant's recording sessions were scheduled, and they were emailed links to an electronic enrolment and consent form hosted on Qualtrics within 72 hours of the first session. This was to minimize the unnecessary collection of data from individuals who agreed to attend but subsequently decided not to participate. Upon completion of the recordings, participants were compensated for their time with e-vouchers redeemable in several shops. For Day participants, these comprised £20 for three sessions (Day 1) and £60 for three sessions (Day 2) to encourage completion of both days. Week participants received £40 for three sessions.

### Data Collection Schedule

Participants in *Week* were scheduled for recording on a Monday, Wednesday and Friday at the same time each day, to minimize within-day variability [41]. Participants in *Day* were scheduled for recording starting between 08:00 and 10:00, 13:00 and 15:00 and 17:00 and 19:00). A minimum time between sessions of 3.5 hours was maintained to maximize the likelihood of measuring differences in speech with time of day. The same participants were scheduled to return for a second day of recording at least eight weeks later. Day 2 of recording was scheduled for the same day of the week as Day 1, and session times were scheduled at the same times as Day 1.

### Recording Session Procedure

At each participant's first session, researchers explained the recording procedure and those who had not already done so in advance of the session completed their enrolment and consent. The forms collected basic sociodemographic data, height (as a proxy of larynx length), information on the participants' voice use habits in the preceding three months, and their level of English, in the case of non-native speakers.

Prior to beginning the study, the project team discussed the clearest and most consistent way to instruct participants. Our aim was to make participants feel as comfortable as possible and encourage natural speech and reproducible positioning during recording. The team fed back to each other as data collection progressed on any difficulties in this regard, and ways to improve participant instruction.
At the start of every recording session, participants were also asked to complete a pre-recording questionnaire on Qualtrics, that collected the times at which participants woke up and got out of bed, when they last ate and drank any liquid, the extent of their voice use that day, how much sleep they had the previous night and if they were experiencing any minor health issues that could affect their voice (Multimedia Appendix 2). The pre-recording questionnaire also included the Pick-A-Mood tool [46]. Participants were also offered a drink of water at the start of each session; we recorded if they took this.

Participants were seated comfortably as possible on an office chair at a desk. Their speech was recorded with an Audio Technica 2020USB+ condenser microphone on a shock mount fitted to a Rylock foam pop filter on a tabletop stand (Figure 1). The microphone was operated using Audacity open-source software running on a Dell Latitude 7440 Laptop (i5 core, 16GB RAM) running Windows 11. The microphone gain was set to a fixed value at the start of every session to maximize the signal-to-noise ratio while avoiding clipping. Participants were positioned 30cm from the condenser microphone, the distance at which the device's frequency response is specified. The chair's height and left-right position were adjusted so that the participant's mouth was level with the pop filter and centered on the microphone. Participants were reminded not to move their chair during session. The participant and set-up was surrounded by acoustic absorbing foam and textiles.

Figure 1. Recording set-up from the side (left) and the front (right).

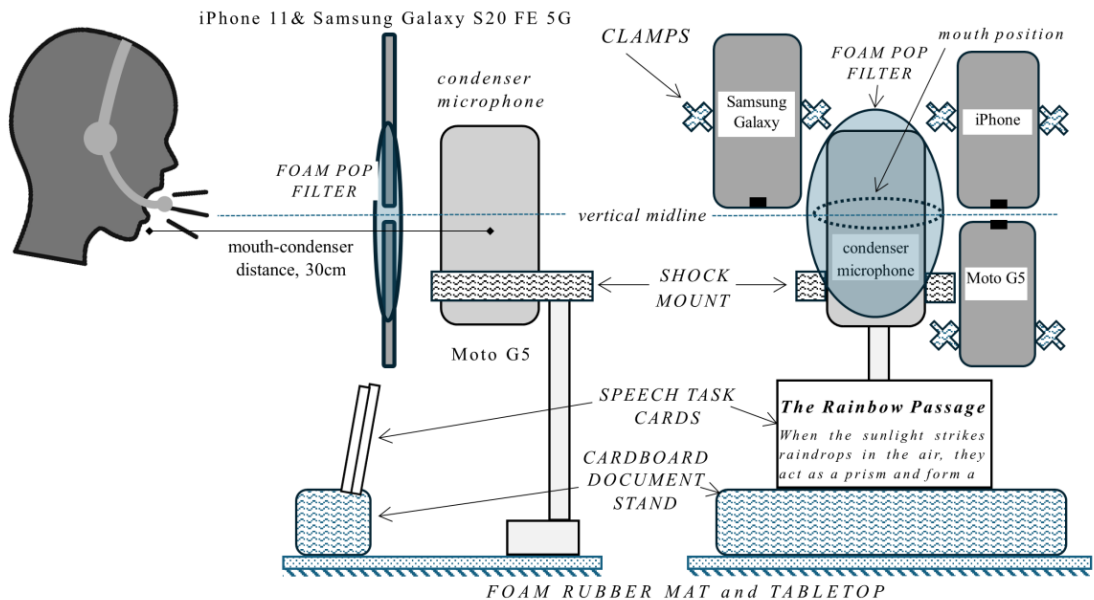

We positioned three smartphones (iPhone 11 (released 2019), Samsung Galaxy S20 FE 5G (released 2020), Motorola G5 (released 2017)) directly adjacent to and in the plane of the pop filter with their microphones positioned on the estimated vertical midline of the condenser microphone. These positions were fixed through all recordings and were comparable to if the participant held their phone in front of them as if in a video call [7]. Smartphone positioning was checked before each session.

Participant also wore a budget consumer office headset (Plantronics Blackwire 3220). Headset microphones combined with a computer or mobile device are an additional potential device of remote speech capture and, as such, we wanted to measure how they might capture within-individual speech variability compared to other devices. Headsets are also recommended by the ASLHA Panel as the microphone-mouth distance can be fixed for the duration of a recording [36]. Our headset was operated using Audacity run on a MacBook Air (Intel Core i5, 16GB RAM), again using a gain level fixed over all participants and sessions. Participants were instructed to position the headset microphone two finger widths from their cheek and to one side of their mouth, using a mirror as needed. The supervising researcher checked headset microphone positioning prior to recording.

Before commencing the elicitation tasks, the participants were instructed to complete them at their own pace and to speak at a natural volume and pace. They were also instructed to switch their phones off or into flight mode or leave them outside the recording room to prevent interference with the recordings.

At the beginning and end of each recording session, as well as between each exercise, the researcher running the session played an audio tone (an alarm tone on their mobile) to prompt the participant to proceed with the next speech task and to aid the manual separation of the tasks into individual audio files following the session.

Following the completion of the speech tasks, the researcher assisted the participant in removing the headset and stopped each recording device. Participants were thanked for their time and reminded of their next recording session appointment, where applicable. At the end of each participant's final recording session, researchers asked participants to consider completing a post-participation questionnaire. Following their departure, the project team promptly emailed participants a link to the questionnaire and codes for shopping e-vouchers, compensating them for their time.

### Speech elicitation tasks

Researchers provided participants with a varying combination of speech elicitation tasks in each session (Figure 2). Our choices balanced collecting several types of speech that when combined provide a variety of health-related indicators, and sufficient amounts of each, with participant burden and acceptance. A protocol with too many tasks, long recording sessions and/or the elicitation of speech with personal content could put off potential participants and result in failures to complete all scheduled sessions.

Figure 2. Speech elicitation overview. Our protocol elicited non-practiced long scripted speech in each session, plus practiced short and long readings, except in Session 1. Participants described a different picture and performed held vowels in each session. The task order was varied between participants and between sessions.

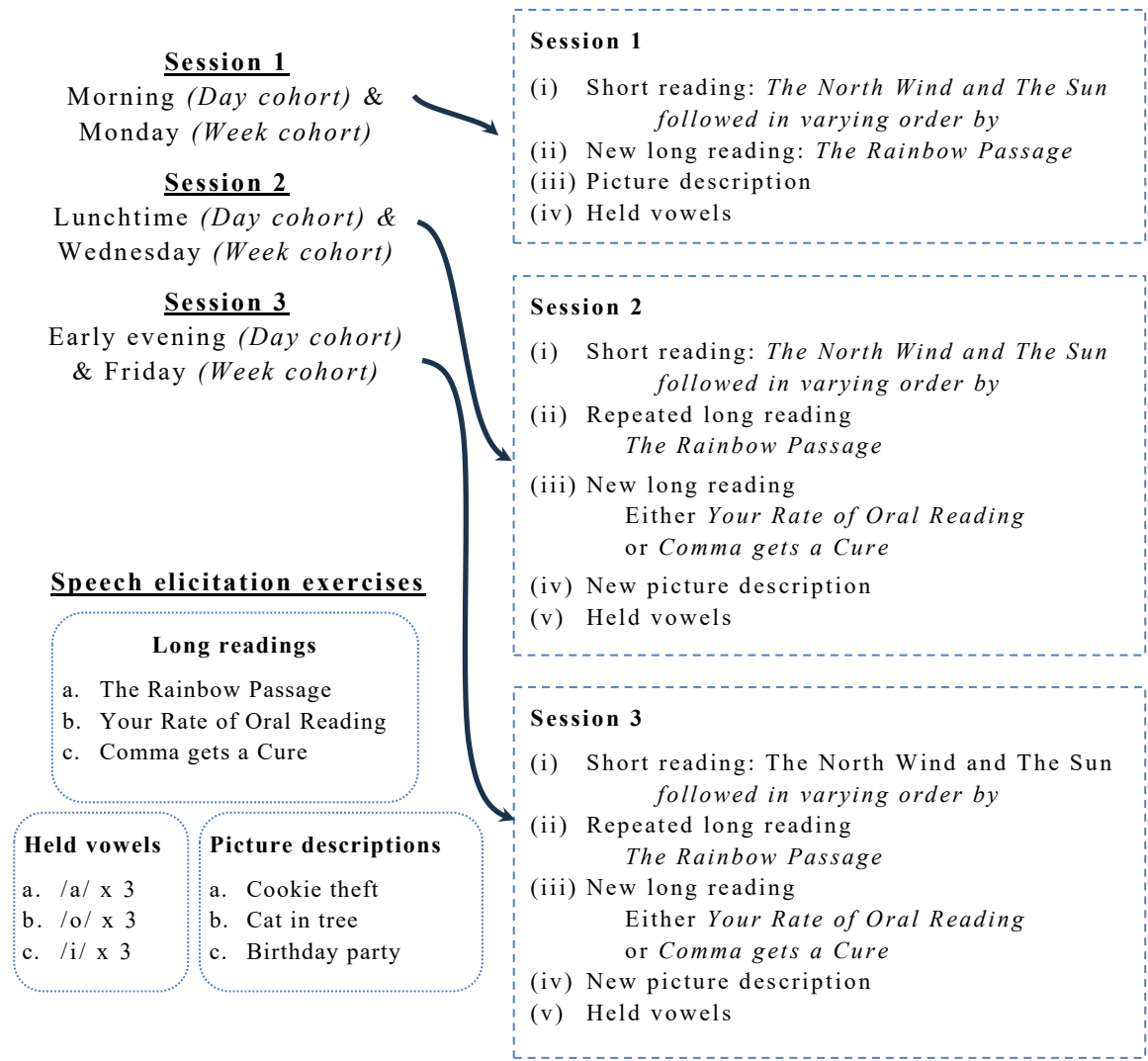

Session 1 began with a short, simple reading, The North Wind and the Sun [47] as a form of warm-up exercise to help participants feel comfortable and settled before beginning the other tasks that would be the focus of our analyses. This was followed by a longer reading, The Rainbow Passage (long version) [48]; a timed picture description (up to two minutes) and three repetitions each of three held vowels, /a/, /o/ and /i/. In Sessions 2 and 3, participants completed the two readings from the first session and the held vowels and an additional long reading in each, one of Your Rate of Oral Reading [49] and Comma Gets a Cure [50]. They also completed a new picture description in each of Session 2 and 3. The elicitation task order was varied between sessions of each participant and between participants to avoid introducing systematic biases with specific tasks.

The scripted tasks provided standardized linguistic content. Repetitions of the North Wind and the Sun and The Rainbow Passage enable direct comparison of paralinguistic features for the same speech between sessions, though these repeated recordings will also be affected by practice effects. Recordings of Your Rate of Oral Reading and Comma Gets a Cure provided set linguistic content that was not subject to practice effects in the Week study and in Day 1 of the Day study, as they were new to the participant.

We selected Your Rate of Oral Reading and Comma gets a Cure as along with the Rainbow Passage, the three readings have a similar lexical and linguistic complexity and length, combined with a similar phonetic balance literature [51, 52]. Therefore, we deemed them suitable for quantifying speech variability between sessions while avoiding practice effects.

The Rainbow Passage and Your Rate of Oral Reading were selected as factual texts rather than stories to minimize the likelihood of participants using a 'story telling' voice and, therefore, maximize the likelihood of them speaking in their natural voices. This choice was informed by our observations in the mHealth study, Remote Assessment of Disease and Relapse – Major Depressive Disorder (RADAR-MDD) [53], where participants tended to use emphasis and be expressive in reading a story. Our choice of Comma gets a Cure was a compromise; it is a story but has desirable lexical and phonetic characteristics that have been well documented in the literature [51, 52].

Picture description tasks provided spontaneous speech. We used three images: The Cookie Theft (original version), The Cat in the Tree, and The Birthday Cake [54]. All pictures are black and white designs, depicting a simple story situation with a central focus and interacting elements. Typically used in speech assessment in neurodegenerative disorders, e.g., Alzheimer's, to investigate cognitive characteristics via the linguistic content of an individual's speech [55], picture descriptions also have value in paralinguistic analysis [56].

Held vowel sounds provided standardized acoustic signals without any lexical, structural or linguistic effects to account for, suitable for measurement of perturbation and quality measures [57, 58]. The choice of elicitation tasks was advantageous from a data privacy perspective as they did not elicit the disclosure of personal information.

### Data Quality Control Checks, Storage and Preparation

After each recording session, all audio files were named with the format ParticipantID_Device_Day_Session. They were then uploaded to a secure Microsoft SharePoint site maintained by King's College London, accessible only by project staff.

The researcher running the session also completed a data quality control log which detailed (i) the start time of each session, using the timestamp on the audio files, (ii) if the participant drank any water during the session, (iii) any interruptions or participant behavior that could affect the recording content or quality, e.g. the participant moving their chair and subsequent chair repositioning, (iv) any extraneous noise during the session, (vi) any issues completing the vowel task, (v) any participant difficulties completing the tasks, and (vii) and any other event or observation not covered by the other fields that could affect the recording.
The researcher then checked that (i) all audio files were uploaded into the correct participant and session folders, (ii) each file contained recordings of the correct speaker, and (iii) all tasks were completed in the stated order. The researcher also noted any additional audible issues in the data not previously captured in the quality control log.

Recordings of individual elicitation tasks were then separated into individual files using Audacity. File names were appended to include which task they contained with the convention ParticipantID_Device_Day_Session_Task.

### Preliminary feature extraction

We extracted 14 exemplar features from condenser microphone recordings of the Rainbow Passage. These features were chosen as they are commonly used in speech-health research, representing timing and fluency characteristics and the speech production subsystems of respiration, phonation and articulation (Table 2).

Table 2. Speech features extracted from the recordings to generate normative.

| Feature | | Description |
|---|---|---|
| **Timing and fluency** | | |
| | duration, s | length of recording |
| | speaking rate, syllables $s^{-1}$ | total syllables divided by duration |
| | articulation rate, syllables $s^{-1}$ | total syllables divided by total speaking time |
| | pause rate, $s^{-1}$ | total pauses divided by duration |
| **Respiration** | | |
| | intensity (mean), dB | loudness of speech signal |
| **Phonation** | | |
| | pitch (mean), Hz | auditory perceived tone |
| | pitch (std deviation), semitones | standard deviation of pitch |
| | harmonic to noise ratio (mean), dB | extent to which harmonic structures are affected by noise |
| | spectral slope (mean) | gradient of the voiced spectrum |

|  | cepstral peak prominence (mean), dB | amplitude of cepstral peak, relative to a regression line through the cepstrum |
|---|---|---|
| **Articulatory** | | |
|  | 1st formant frequency (mean), Hz | 1st resonant frequency of the vocal tract |
|  | 2nd formant frequency (mean), Hz | 2nd resonant frequency of the vocal tract |
|  | gravity (mean), Hz | center frequency of the narrow band spectrum |
|  | deviation (mean), Hz | spread of frequencies around the spectral gravity |

Timing and fluency features have previously been demonstrated to contain important clinical information for conditions including depression [7], [59], ALS [60] and Parkinson's disease [61]. Respiration and phonation features are widely used in speech-based mental health analysis [7], [9], [10]. Articulation features have been included as they indicate changes in speech intelligibility and speech-motor control and have been proposed as markers for a variety of health conditions [9-11, 62].

To extract these features, we first used Parselmouth [63] to convert all audio files to single-channel 16kHz Waveform Audio File Format (WAV) files with 16-bit resolution. Our acoustic features were extracted at two levels: (1) suprasegmentally: calculated over the entire reading, and (2) for individual occurrences of open /a/ vowels of at least 50ms duration from the Rainbow Passage. For the /a/ vowels, we extracted the features per identified instance of the vowel and calculated the mean per recording over all instances. We provide suprasegmental acoustic features, as this is a common approach in paralinguistic analyses [64]. Extraction purely from /a/ vowels, in contrast, provides more granular, controlled acoustic measures of speech. The use of the open /a/ vowel has been recommended for more reliable extraction of voice quality measures [65].

As a more realistic and affordable approach towards clinical research, we implemented an automated approach to identify instances of /a/ in our files. First, we transcribed our files using the Open AI whisper-base.en model [66], an open-source Automatic Speech Recognition (ASR) tool, which has been demonstrated, in independent testing, to have an average word error rate of 12.8% calculated over nine different ASR test sets [67]. We then performed a forced alignment of the resulting transcripts utilizing the Montreal Forced Aligner (MFA) [68] and English MFA acoustic model V2.0.0a . After identifying the vowels in the phonetic alignment, we extracted the features per identified vowel, then took the per-participant, per session mean of these features to form our final representation. We performed spot-checks of the accuracy of these alignments, dictated by timing and budgetary constraints.

Features were extracted using Parselmouth [63], an open-source Python library that enables the use of Praat, a software package for speech analysis [69]. Speech timing features are extracted using intensity thresholds [70]. All prosodic, phonation and articulatory measures were extracted using default Praat settings, except for the extraction of F0, which followed the two-step approach recommended in [59] and Cepstral Peak Prominence, which followed settings recommended in [71]. All code is available on request.

## Summary

Our protocol is unique (Table 1): it collects using multiple microphone types in a controlled environment to control for and minimize variability attributable to hardware, set-up, and acoustic conditions. The speech elicitation prompts enable the collection of acoustically rich and varied content while (i) containing a core amount of fixed phonetic content to enable comparable analyses and (ii) introducing new readings in each session to minimize potentially confounding practice effects. We collated a list of factors we considered in designing our protocol that may be used as a framework by other researchers designing speech collection protocols (Multimedia Appendix 1).

## Results

Recruitment began on June 5th, 2023. Pre-enrolment screening to exclude any hearing, speaking, neurological or mental health disorders that might affect their speech was completed by 141 participants (Figure 3). In total, 28 and 26 participants enrolled in the Day and Week studies, respectively (Table 3). One participant in Week completed two of the three recording sessions due to illness (Figure 3). Day 1 recordings began on June 14th, 2023, and were completed on August 10, 2023. Day 2 recording began on August 9th, 2023. and were completed on October 5, 2023. Week recordings commenced on June 19, 2023, and were completed on October 6, 2023.

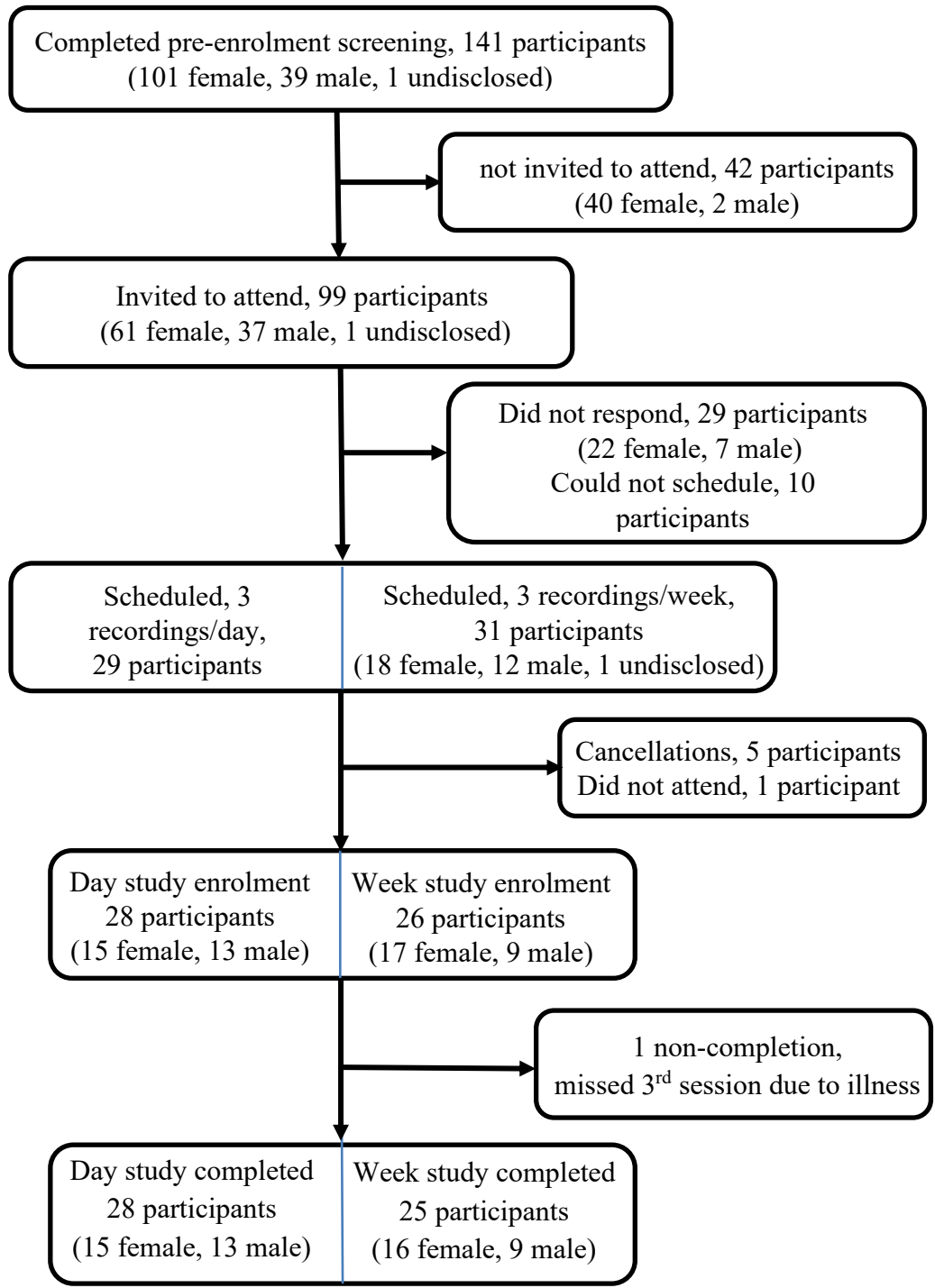

Figure 3. STROBE flowchart describing participant recruitment, enrolment and completion. Pre-enrolment was completed via Qualtrics, email and face-to-face.

Table 3. Participant characteristics.

| Characteristic | Day | Week |
|---|---|---|
| **Sex** | | |
| female | 15 | 17 |
| male | 13 | 9 |
| **Age (years)** | | |
| median (quartile 1, quartile 3) | 26 (23, 34) | 29 (24, 34) |
| **Height (m)** | | |
| median (quartile 1, quartile 3) | 1.70 (1.63, 1.79) | 1.71 (1.63, 1.78) |
| **Ethnicity** | | |
| White, UK & Ireland | 14 | 9 |
| White, other | 3 | 5 |
| Asian/Asian British (Indian, Bangladeshi & Chinese) | 5 | 7 |
| Black/African/Caribbean/Black | 1 | 1 |
| British - Caribbean | 0 | 1 |
| Arab | 0 | 1 |
| Mixed/Multiple ethnic groups | 4 | 1 |
| Other Ethnic group | 1 | 2 |
| **First Language** | | |
| native | 24 | 17 |
| non-native[a] | 4 | 9 |
| **Voice use 3 mos. prior** | | |
| low | 1 | 2 |
| intermittent | 7 | 6 |
| regular | 19 | 15 |
| high | 1 | 3 |
| **Minor health issues** | | |
| allergies | 2 | 4 |
| sinusitis | 1 | 0 |
| Acid reflux | 1 | 1 |

[a] All B2 level or above per the Common European Framework of Reference for Languages, www.coe.int.

In Day, the median recording start times for the morning sessions was 9.12 and 9:11 for Day 1 and Day 2, respectively (Multimedia Appendix 3 and 4). The median afternoon and evening recording start times for both days were 14:05 and 18:04. In Week, the most common recording slots were 10:00-11:00 and 12:00-13:00, with 5 participants each (Multimedia Appendix 5). Recording times for each participant were consistent across the Monday, Wednesday and Friday sessions, with differences in start times all less than 30 minutes (median: 13 minutes, range: 3-22 minutes).

In total, the study comprised 245 recording sessions and produced 1,225 audio files from five recording devices, totaling 169 GB of data. Using Audacity, we separated the readings of *The Rainbow Passage* from the condenser microphone and extracted our 14 exemplar speech features using the methodology previously outlined. These values are provided for Day and Week participant groups (Table 4). We present features captured with the condenser microphone only as our benchmark device, as exemplar values that are not subject to any pre-processing that could erroneously affect the values extracted.

Table 4. Normative values, median (1st quartile, 3rd quartile) for a set of exemplar timing, prosody, voice quality, articulation and spectral features.[a]

| L[b] | Week | Day, Day 1 | Day, Day 2 |
|---|---|---|---|
|  | n = 26 | n = 28 | n = 28 |
|  |  |  |  |
| **duration, s** | | | |
| S | 122 (110, 136) | 114 (104, 126) | 113 (103, 123) |
| **Speaking rate, syllables s$^{-1}$** | | | |
| S | 3.69 (3.40, 3.99) | 3.72 (3.53, 4.06) | 3.72 (3.49, 4.00) |
| **Articulation rate, syllables s$^{-1}$** | | | |
| S | 4.63 (4.32, 4.91) | 4.65 (4.47, 4.87) | 4.65 (4.48, 4.91) |
| **Pause rate s$^{-1}$** | | | |
| S | 0.233 (0.205, 0.264) | 0.215 (0.191, 0.255) | 0.214 (0.187, 0.251) |
| **Pitch (mean), Hz** | | | |
| S | 187 (120, 202) | 146 (111, 194) | 154 (112, 192) |
| a | 183 (124, 203) | 150 (114, 189) | 157 (115, 196) |
| **Pitch (std deviation), Hz** | | | |
| S | 2.91 (2.54, 3.57) | 2.78 (2.35, 3.79) | 2.89 (2.41, 3.69) |
| a | 0.37 (0.26, 0.59) | 0.38 (0.22, 0.56) | 0.35 (0.21, 0.58) |
| **Intensity, dB** | | | |
| S | 68.7 (67.1, 70.1) | 68.2 (66.7, 69.8) | 68.5 (66.8, 69.9) |
| a | 72.7 (70.8, 74.5) | 72.6 (70.2, 74.6) | 73.0 (70.7, 74.7) |
| **Harmonic to noise ratio, dB** | | | |
| S | 10.43 (7.28, 12.02) | 8.33 (6.39, 9.85) | 8.34 (6.69, 9.97) |
| a | 8.57 (4.62, 10.61) | 6.44 (4.38, 7.97) | 5.58 (3.72, 8.33) |
| **Spectral slope** | | | |
| S | -17.0 (-18.6, -15.6) | -16.4 (-17.7, -15.4) | -16.4 (-18.0, -15.1) |
| a | -20.5 (-22.0, -19.1) | -19.8 (-21.6, -18.7) | -19.7 (-22.1, -18.7) |
| **Cepstral peak prominence, dB** | | | |
| S | 10.14 (9.56, 10.74) | 9.96 (9.43, 10.69) | 10.17 (9.36, 10.77) |
| a | 13.91 (12.84, 15.03) | 13.26 (11.76. 15.03) | 13.50 (11.91, 14.82) |
| **First formant, Hz** | | | |
| S | 477 (449, 504) | 482 (454, 507) | 475 (450, 505) |
| a | 648 (577, 698) | 639 (580, 674) | 628 (571, 678) |
| **Second formant, Hz ×10³** | | | |
| S | 1.65 (1.56, 1.72) | 1.57 (1.49, 1.64) | 1.57 (1.48, 1.63) |
| a | 1.33 (1.25, 1.43) | 1.26 (1.17, 1.36) | 1.25 (1.17, 1.37) |
| **Spectral gravity, Hz** | | | |
| S | 417 (362, 465) | 453 (388, 487) | 433 (367, 497) |
| a | 613 (511, 687) | 651 (564, 706) | 621 (544, 696) |
| **Spectral deviation, Hz** | | | |
| S | 330 (286, 389) | 363 (324, 398) | 357 (320, 393) |
| a | 361 (310, 426) | 370 (342, 414) | 362 (331, 407) |

[a] Feature definitions are provided in Table 2.
[b] Feature extraction level (L): features are extracted suprasegmentally (*S*) and/or from automatically identified /a/ vowels (*a*) in readings of the Rainbow Passage recorded with a condenser microphone that did not apply any pre-processing.

The focus of this paper is methodology development. Therefore, an analysis of within-individual speech variation and the ability of different devices to capture this variation is beyond the scope of this paper; it will be reported in future publications.

## Discussion

We developed a protocol to record repeated speech samples in the same individuals over time. The metadata reporting, scheduling, device choices, elicitation tasks, data storage and preparation and feature extraction provide an adaptable template for other researchers collecting repeated speech samples.

Our specific research focus was to gain insights into speech variation over the course of a single day and week while controlling for practice effects. The protocol is unique in studies exploring within and between-speaker variability in a non-pathological population in the variety of speech captured and the number and type of recording devices. This allows us to observe how within-individual variability is captured by mobile devices. Analysis of these aspects will be presented in future work.

The protocol also enabled us to generate a small but well-described dataset of normative values of 14 exemplar features commonly used in speech-health research. The insights resulting from this work provide us with a foundation for the design of future data collection and interpretation in clinical cohorts.

### Limitations & lessons learned

The design and implementation of this protocol provided insights that will inform methodology of future studies.

#### Protocol Development

The design of this protocol was made challenging by the absence of suitable established collection and reporting protocols [34]. Discipline silos are a core challenge in speech-based health assessment literature that hinders protocol development and reporting. There is a lack of teams integrating clinical-facing researchers who collect data and researchers who process and analyze the data, who are typically from engineering or computer science backgrounds. This can lead to gaps in the collection and reporting of speaker factors and methodological choices that can influence the measurement of recording speech. Consistent reporting of the effects of speaker, recording and processing factors is urgently required to aid the development of robust speech collection protocols and processing pipelines [10, 72] and to inform the statistical design of speech studies in clinical cohorts [73].

### Resource requirement

Though participant numbers were small (n=54), the resources required to implement all steps of the protocol – recruitment, data collection, pre-processing of audio files, and feature extraction – were substantial. Data collection and audio preprocessing were particularly labor-intensive. Our 245 recording sessions extended from 8:00 am to 7:30 pm. We preferred to run these sessions with two researchers present to help minimize the likelihood of errors, though this was often not logistically feasible. Regarding pre-processing, we estimate that splitting the 1,225 audio files into their individual tasks required close to 720 hours of researcher time. This highlights the need for more efficient recording and annotation techniques to recruit large, well-powered studies.

One way to increase dataset size and minimize researcher burden when implementing a similar protocol in the future could be to collect data remotely using personal computers or smart devices using collection platforms such as RADAR-base [74]. Such a solution does not require researcher time to run the recording sessions, and apps can be easily designed to record different speech elicitation activities individually, saving manual segmentation time. However, remote studies are more likely to result in missing data, incorrectly completed tasks and more variable data quality [53, 74].

Participant non-compliance, particularly in clinical cohorts, is a further concern in remote studies. Pierce et al. reported high adherence of 92% of their healthy participants to the prescribed recording times over seven days [41]. Over collection intervals of up to 18 months, we observed clinical cohort completion rates of 50% (IQR: 21-74%) and 41% (IQR: 13-67%) for the scripted and free-speaking speech tasks, respectively, in RADAR-MDD, where speech was one of more than 10 data streams. Within the sparse longitudinal literature, the Voiceome study is a further example, where only 21% of participants completed two or more recordings [37]. Therefore, there is a need to understand participant motivation and functionality concerns in mobile data collection.

### Recruitment balance

Before beginning recording, we aimed to recruit a 50-50 balance of sex at birth. However, we quickly learned that this required a concerted effort to achieve in the fixed time that we had to complete our work dictated by funder requirements. In total, 101 women completed pre-enrolment forms versus 39 men, which was only achieved following specific appeals for male participants. Our final overall cohort comprised 22 men and 32 women. While not 50-50, this is more balanced than the 75-25 female-male balance of the clinical speech cohort recruited in RADAR-MDD, that was attributed to the greater reported incidence of depression in women [7], [53]. We did achieve good attendance once participants enrolled, with only one participant in 54 missing one session due to illness. This highlights that participants were engaged and willing to complete the speech tasks.

### Recruitment feasibility in clinical cohorts
When implementing our protocol, we benefited from the large pool of potential 'healthy' volunteers in our institution. Clinical inclusion criteria could shrink the recruitment pool, and staff and students may be more reluctant to volunteer if it requires disclosure of a diagnosed mental health disorder. Therefore, it remains to be seen if a clinical cohort, such as participants with MDD, could be recruited for the same protocol recording in a controlled environment, given the need for set recording times and days for 3-6 sessions.

In separate research in a clinical cohort, we have observed that the choice of speech elicitation activity is also important for participant and patient engagement in the context of future mobile speech monitoring applications [38]. Fixed, repeated tasks increase the risk of disengagement; for example, we received participant feedback in RADAR-MDD that repeating the same reading every two weeks for up to two years became tedious. Recruitment in the Voiceome study was high, but data contribution rates were low, and the lack of engagement was not discussed [37].

### Metadata collection
A range of speaker-specific factors dictate changes in speech; therefore, the collection of personal data is essential in speech-health studies as such factors may relate to selection, information or confounding biases. The collection of such information is a balancing act of analytical goals versus (1) ethical considerations that dictate any personal information collected should only relate to what is needed for obtaining meaningful results, (2) participant acceptance and recruitment feasibility, as studies collecting more personal and sensitive information, which may also increase the participation time, maybe more challenging to recruit and (3) logistical considerations, depending on the time and resources available to complete data collection.

We had ethical, participant acceptability and logistical factors in mind when deciding what information to collect in our protocol (Multimedia Appendix 2). Information that we did not collect but would recommend others consider include (i) caffeine and alcohol intake prior to recording [75, 76], (ii) menstrual cycle phase at the time of recording and whether female speakers are menopausal [12, 13], (iii) participant mood using a clinically validated tool.

As this protocol was for a pilot study, we did not consider getting feedback on metadata collection through Patient Public Involvement (PPI) work. This should, however, be a core consideration when utilizing the underlying methodology in future studies.

### Equipment setup

Our set-up had two limitations with implications for speech measurement precision. Firstly, it was possible for participants to move the position of the office chair on which they were seated during recording as it had wheels and was rotatable. This was a trade-off, as with the chair's height adjustment feature, participants could be easily centered on the microphone set-up per our protocol. We mitigated this risk by observing participants during recording and making gentle reminders not move the chair and in rare cases, repositioning the participants. However, participant movement could not be completely excluded.

Secondly, there was a limit on how close participants could position their mouths from the microphone, depending on their BMI, as the condenser microphone was set back from the table edge in a fixed position for the study to minimize adjustment of the set-up and fully surround it by the acoustic foam enclosure. This had the potential to result in deviations from mouth-microphone distance in our protocol. This issue could be mitigated by positioning the microphone closer to the desk edge, combined with an extension of the acoustic foam to surround the participant and microphone more fully.

Additionally, early in the study, we occasionally observed small amounts of audible interference on recordings from mobile phones and on rare occasions, phone alert tones and incoming calls. We subsequently requested that participants switch their mobile devices off, place them in flight mode or leave them outside the recording room during sessions. We later began to set our study phones in flight mode, after occasional, new observations of interference in sessions where interference from the participants' phone could be excluded.

### Feature Extraction

Our choice and specification of features to report represented a considerable challenge when developing the protocol. To the best of the authors' knowledge, there is no agreed minimal benchmark feature set in the literature for such a purpose. Additionally, the perturbation and quality measures typically reported in the voice disorder literature [57, 58] are limited; they do not adequately capture all the vocal effects associated with neurological and mental health conditions.

Meanwhile, pre-defined multivariate feature sets such as the extended Geneva Minimalistic Acoustic Parameter Set (eGeMAPs), or the Computational Paralinguistics ChallengE Set (ComParE), available in the openSMILE toolkit [77, 78] were not designed for health-assessments. For example, these feature sets do not contain specific timing and fluency measures such as pause rate, a widely utilized feature in the ALS and depression literature [7, 8, 60]. A similar feature set to ours is published in [80], but it contains jitter and shimmer measurements, which have limited utility when extracted from connected speech [65].

An additional challenge is that many commonly reported features are not uniformly defined or extracted by different extraction tools. For example, Lenain et al., compared vocal jitter across three toolboxes and only obtained weak correlations between the different implementations [81].

We used Praat as it is arguably more widely used in speech pathology and phonetics research. A weakness of Praat we observed, however, relates to the number of settings associated with extracting each feature; finding guidance on preferred values for these settings is difficult. We also observed that default values were not ideal in certain circumstances. For example, when testing the pitch feature extraction code, we observed that the default pitch ceiling value of 500 Hz could result in false pitch readings of over 300 Hz, well outside of expected ranges for this feature.

A challenge relating to extracting features over specific vowels is reliance on third-party automatic speech recognition (ASR) and forced alignment tools. Our choice of Whisper and the Montreal Forced Aligner was to allow us to extract normative feature values from a processing pipeline comprising standard, open-source, well-established tools. Due to resource constraints, we were limited to spot-checks of alignments. However, in subsequent work using this dataset, we have observed differences in timing features extracted using word boundaries estimated from transcripts generated using different ASRs [26]. Further work, including manual verbatim and phonetic transcriptions, is required to explore the effects of ASR performance and alignment accuracy on the quality and reliability of transcripts and vowel locations [82].

*Data utility*
Though the core motivation in developing our protocol was within-individual speech variation within one day and one week, towards longitudinal assessments of health, the resulting data has broader utility in speech research and therefore represents value for funding. This is important to consider in study design given the large resources needed to generate speech corpora.

We have begun using the data to benchmark different speech technologies (e.g., Automatic Speech Recognition) and quantify associated variability in the feature extraction pipeline [26]. We have also demonstrated practice effects in repeated readings [83]. Further utility is gained from recording over multiple devices and using different elicitation methods, allowing us to assess variability in speech features according to these key methodological choices. It is vital to characterize such variation in speech over repeated speech samples, to identify and develop reliable speech markers pipelines for clinical research and practice. Finally, we are also preparing to make the datasets accessible to other non-profit researchers, enabling other investigations.

## Conclusions

Within the speech-based health assessment literature, core methodological details and speaker characteristics are often under-reported and/or the rationale for choices not explained. With this in mind, we have described a protocol for collecting non-pathological repeated speech samples. The core methodological aspects of this protocol cover design and reporting decisions are relevant for researchers collecting longitudinal data for speech and language biomarker research. As a harmonization step, we encourage other researchers to adopt these aspects in their own projects, thereby adding replicability and, ultimately, the translation of speech and language biomarkers into clinical research and practice.

## Acknowledgements

We thank our participants for their valuable support and the King's College London Department of Psychology for use of their test rooms for recording.
This project was funded by a combination of an Engineering and Physical Sciences Research Council and UK Acoustics Network Plus grant (reference: EP/V007866/1) and an IPEM Innovation Award (reference: N/A).

This research is part funded by the National Institute for Health and Care Research (NIHR) Maudsley Biomedical Research Centre (BRC). The views expressed are those of the author(s) and not necessarily those of the NIHR or the Department of Health and Social Care.

ZR and CL were supported by King's Undergraduate Research Fellowships. TP was supported by a Wellcome Trust Summer Internship.

MIT LL Disclaimer: Approved for public release. Distribution is unlimited. This material is based upon work supported by the Under Secretary of Defense for Research and Engineering under Air Force Contract No. FA8702-15-D-0001. Any opinions, findings, conclusions or recommendations expressed in this material are those of the author(s) and do not necessarily reflect the views of the Under Secretary of Defense for Research and Engineering.

NC & JuD: conception, design, acquisition, analysis, data interpretation, manuscript drafting and editing. LW: design, acquisition, analysis, manuscript drafting and editing. EC: conception, analysis, manuscript editing. CL, ZR and TP: acquisition and analysis, FM: Conception, design, manuscript editing. JoD & RD: Conception, manuscript editing. TQ: Conception, data interpretation, manuscript editing

## Conflicts of Interest

NC is a consultant to thymia ltd. The authors have no other disclosures to make, financial or non-financial.

## Abbreviations

ASR: automatic speech recognition
ALS: amyotrophic lateral sclerosis

ComParE: computational paralinguistics challengE set
eGeMAPS: extended Geneva minimalistic acoustic parameter set
MDD: major depressive disorder
MFA: Montreal forced aligner
RADAR: remote assessment of disease and relapse
WAV: Waveform Audio File Format

## Multimedia Appendix 1 (separate file also uploaded)

Checklist of methodological aspects for consideration in protocol design and reporting

| Aspect | | Core considerations |
|---|---|---|
| **Participants** | | |
| | Input and Feedback | Active involvement of patients and the public in clinical speech analytics development is critical to ensure technologies meet real-world needs. Patient and Public Involvement and Engagement (PPI-E) should be considered from the development stage of any speech collection project and can include the following aspects:<br>- project descriptions, and consent documents<br>- recording device and set-up<br>- acceptability of speech elicitation prompts<br>- participant instructions<br>- choice of speech measures<br>- choice of clinical outcome and analysis |
| | Eligibility criteria | Inclusion criteria and their implications for answering the research question in mind, including the presence of confounding factors, and recruitment feasibility:<br>- sociodemographic factors, e.g., gender, sex, age, education level, language, ethnicity<br>- vocal tract and hearing disorders<br>- respiratory conditions<br>- mental health disorders<br>- neurological disorders<br>- lifestyle factors, e.g., smoking status and alcohol consumption |
| | Recruitment | Recruiting a planned sample size within a defined time frame is a key bottleneck in speech research, particularly if a specific balance in, e.g., gender or age is sought. Key aspect to consider include:<br>- sociodemographic and socioeconomic biases<br>- clinical vs general populations |

|  |  | - whether partnerships with advocacy groups, clinical centers or related organizations are needed<br>- Time limitations due to funder requirements |
|---|---|---|
| **Data Collection** |  | Changes in speech are dictated by a range of speaker-specific factors and recording and analysis choices. It is important to collect and report information relating to these factors as they may relate to selection, information or confounding biases. |
|  | Metadata | The collection and reporting of participant characteristics that may be potential confounders |
|  | Clinical assessments | How core clinical outcomes will be assessed. We recommend the use of validated scales and tests, where possible. Considerations include<br>- whether tests are clinician vs self-reported<br>- time required to complete assessment and the associated participant burden |
|  | Recording devices | Effects of recording device, environment and time; these can all cause subtle changes in speech measurement. Aspects to consider and report include:<br>- mobile versus non-mobile devices<br>- omni-directional or uni-directional microphones<br>- ambient noise in the recording location<br>- affordability and accessibility<br>- device-specific signal processing<br>- gain settings |
|  | Recording set-up and environment | Consistent conditions and device-speaker positioning are ideal and relevant factors include:<br>- recording device positioning requirements<br>- ambient noise and room acoustics<br>- participant comfort<br>- whether participants stand or sit<br>- office furniture features, e.g. adjustable seating<br>- positioning of prompts/reading materials<br>- remote versus in-lab or in-clinic data collection<br>- collection of background noise for reporting of signal-to-noise ratio |
|  | Speech elicitation task | Choosing the optimal task to maximize the likelihood of identifying key correlates, clinical or otherwise. Factors to consider include:<br>- whether voice warm-up and familiarization is needed<br>- practice effects and their implications for analysis<br>- task order and risk of associated bias<br>- task difficulty<br>- task acceptability (established through PPI-E) |

|  |  | - collection Procedure (some tasks, e.g., sustained phonation, may require more detailed instructions than others, this could the validity of the recorded sample) |
|---|---|---|
|  | Participant instructions | Decide strategies for instructing participants during sessions informed by PPI-E, adjusting as necessary during the study. Factors to consider include:<br>- how best to ensure reproducible positioning<br>- participant mental and physical comfort throughout session<br>- how to elicit natural speech<br>- how to provide feedback and encouragement |
|  | Data Quality Log | Any incidents or participant behavior, or deviations from the protocol should be logged, where possible, to aid interpretation of the recordings and features subsequently extracted |
| **Data Processing** |  | There are many ways in which we can digitize and process speech that can all affect the recorded signal and influence analysis. |
|  | Digitization | Digitization dictates the information captured and stored. Factors to consider and report include choice of<br>- sampling rate<br>- bit rate<br>- audio file format |
|  | Data Preparation | Resources required to prepare data for feature extraction may be considerable<br>- remove audio before and after each elicitation task<br>- separate different tasks into individual files, where different elicitation tasks are captured in a single file perform any manual checks required, including when automated methods are used |
|  | Preprocessing | Application and reporting of the use of denoising, dereverberation, signal enhancement, speaker separation, and other similar audio processing tools; typically, these are not explicitly developed for clinical applications and may remove or alter health-related signals in the speech. Use of any such tool must be reported. |
|  | Feature selection | - whether the chosen features are measuring a speech construct related to the clinical outcome<br>- whether interpretable/explainable features are needed |
|  | Feature extraction | Feature extraction methodology is a source of variability, including:<br>- choice of transcription tool (if used)<br>- choice of alignment tool (if used) |

|  |  | - choice of feature extraction software and key settings<br>- level of feature extraction (e.g., suprasegmental vs vowels)<br>- criteria for and, identification and removal of outliers |
|---|---|---|

## Multimedia Appendix 2 (separate file also uploaded)

Pre-recording questionnaire completed by participants at the start of each recording session

1. Are you experiencing any minor health issues today that may affect your voice? e.g., hay fever.
    - Yes – please tell us here
    - No
    - Unsure – please tell us here

2. At what time did you wake up today? Please answer to the nearest 15 minutes.

3. At what time did you get out of bed today? Please answer to the nearest 15 minutes.

4. To the nearest hour, how many hours of sleep did you have last night?

5. Which of the following best describes how you have used your voice so far today?
    - Low Activity
        - I have spoken for less than one hour today
        - I haven't spoken above conversational volume
        - I haven't spoken in a group discussion, been teaching, or given a presentation or equivalent
    - Intermediate
        - I have been talking intermittently to frequently today
        - I have raised my voice above conversational levels for short spells
    - High activity
        - I have been talking for long spells today
        - I have been talking loudly and/or with an expressive voice
        - I have been teaching, have given presentations and/or performances

6. When did you last drink something?
    - I had something to drink when I arrived at the recording session
    - Within the last hour
    - More than 1 hour ago
    - More than 2 hours ago

o   More than 3 hours ago

7. When did you last drink something?
    o   Within the last hour
    o   More than 1 hour ago
    o   More than 2 hours ago
    o   More than 3 hours ago

8. How are you? Select the image number from *Pick-A-Mood* that best describes how you feel at the moment.

9. [on a separate screen] To follow up the picture you chose, which of these best describes how you are feeling at the moment? *This question is optional*
    o   neutral
    o   excited-lively
    o   cheerful-happy
    o   tense-nervous
    o   irritated-annoyed
    o   sad-gloomy
    o   bored-weary
    o   calm-serene
    o   relaxed-carefree

## Multimedia Appendix 3 (separate file also uploaded)

*Day* reported recording times, median (quartile 1, quartile 3). Day 1 and Day 2 were 8-11 weeks apart, on the same weekday.

| Session | Time, Day 1 | Time, Day 2 |
|---|---|---|
| **Morning (S1)** | | |
| | 09:12 (08:43, 09:53) | 09:11 (08:33-09:53) |
| **Afternoon (S2)** | | |
| | 14:05 (13:25-14:41) | 14:05 (13:07-14:42) |
| **Evening (S3)** | | |
| | 18:04 (17:35-18:38) | 18:04 (17:30-18:41) |

## Multimedia Appendix 4 (separate file also uploaded)

Intervals between sessions, median (quartile 1, quartile 3) in minutes for Day 1 and Day 2 in *Day*.

| | Day 1 intervals | Day 2 intervals |
|---|---|---|
| | | |
| **S1-S2** | | |

|  | 291 (287, 292) | 293 (290, 298) |
|---|---|---|
| **S2-S3** | | |
|  | 240 (237, 241) | 240 (238, 241) |
| **S1-S3** | | |
|  | 529 (525, 533) | 534 (530, 538) |

**Multimedia Appendix 5** (separate file also uploaded)

Distribution of recording times in *Week*.

| Recording interval | Number of participants |
|---|---|
|  |  |
| 09:00-10:00 | 1 |
| 10:00-11:00 | 5 |
| 11:00-12:00 | 2 |
| 12:00-13:00 | 5 |
| 13:00-14:00 | 2 |
| 14:00-15:00 | 2 |
| 15:00-16:00 | 3 |
| 16:00-17:00 | 4 |
| 17:00-18:00 | 1 |